\documentclass[journal]{IEEEtran}
\usepackage{cite}
\usepackage{amsmath,amssymb,amsfonts}
\usepackage{algorithmic}
\usepackage{algorithm}
\usepackage{graphicx}
\usepackage{stmaryrd}
\usepackage{xcolor}
\usepackage{array}
\usepackage{commath}
\usepackage{sidecap}
\usepackage{stfloats}
\usepackage{tabularx, boldline}
\usepackage{rotating,booktabs,multirow}
\usepackage{mathtools}
\usepackage{flexisym}
\usepackage{breqn}
\usepackage{url}
\usepackage{adjustbox}
\usepackage{makecell}

\usepackage{textcomp}
\usepackage{amsmath}
\usepackage{amsthm}
\usepackage{txfonts}
\usepackage{graphicx}%,subcaption}
\usepackage{acronym}
\usepackage{tabularx, boldline}
\usepackage{rotating,booktabs,multirow}
\usepackage{enumerate}
\usepackage{enumitem}

\def\E{\mathbb{E}} % Esperanza
\ifCLASSINFOpdf
\else
\fi

\begin{document}

\title{Islanding Detection for Active Distribution Networks Using WaveNet+UNet Classifier}
%
%
% author names and IEEE memberships
% note positions of commas and nonbreaking spaces ( ~ ) LaTeX will not break
% a structure at a ~ so this keeps an author's name from being broken across
% two lines.
% use \thanks{} to gain access to the first footnote area
% a separate \thanks must be used for each paragraph as LaTeX2e's \thanks
% was not built to handle multiple paragraphs
%

\author{Amirhosein~Alizadeh,
        Seyed Fariborz~Zarei,%~\IEEEmembership{Member,~IEEE,}
        and~Mohammadhadi~Shateri%,~\IEEEmembership{Member,~IEEE}% <-this % stops a space
\thanks{A. Alizadeh and S. F. Zarei are with the
 Electrical and Computer Engineering Department, Qom University of Technology, Qom, Iran.

 Corresponding author: S. F. Zarei (e-mail: zarei@qut.ac.ir).}% <-this % stops a space
\thanks{M.Shateri is with the Systems Engineering Department, École de technologie supérieure, Montreal, Canada.}}% <-this % stops a space
\maketitle

% As a general rule, do not put math, special symbols or citations
% in the abstract or keywords.
\begin{abstract}
This paper proposes an AI-based scheme for islanding detection in active distribution networks. By reviewing existing studies, it is clear that there are several gaps in the field to ensure reliable islanding detection, including (i) model complexity and stability concerns, (ii) limited accuracy under noisy conditions, and (iii) limited applicability to systems with different types of resources. Accordingly, this paper proposes a WaveNet classifier reinforced by a denoising U-Net model to address these shortcomings. The proposed scheme has a simple structure due to the use of 1D convolutional layers and incorporates residual connections that significantly enhance the model's generalization. Additionally, the proposed scheme is robust against noisy conditions by incorporating a denoising U-Net model. Furthermore, the model is sufficiently fast using a sliding window time series of 10 milliseconds for detection. Utilizing positive/negative/zero sequence components of voltages, superimposed waveforms, and the rate of change of frequency provides the necessary features to precisely detect the islanding condition. In order to assess the effectiveness of the suggested scheme, over 3k islanding/non-islanding cases were tested, considering different load active/reactive powers values, load switching transients, capacitor bank switching, fault conditions in the main grid, different load quality factors, signal-to-noise ratio levels, and both types of conventional and inverter-based sources.
\end{abstract}

% Note that keywords are not normally used for peerreview papers.

\IEEEpeerreviewmaketitle

\section{Introduction}

\IEEEPARstart{I}{slanding} is a condition where portions of the network Embrace independence from the main grid by autonomously generating and utilizing electricity. ~\cite{RN1}. This presents a significant challenge for distribution system operators when dealing with active distribution networks~\cite{n01}. From an electrical perspective, islanding can lead to voltage and frequency fluctuations, degraded power quality, and potential damage to infrastructure components\cite{mumtaz2023extensive}. On the safety front, islanding poses a considerable risk to line workers who may be unaware of localized energized islands during maintenance tasks\cite{pd4}. Given these implications, it is imperative to detect islanding conditions to uphold the dependability of the system and protect the personnel~\cite{n02}.
%Islanding detection methods encompass diverse categorizations in terms of principles and techniques, including classical schemes, remote schemes, and modern schemes~\cite{RN9}. 
Islanding detection methods encompass diverse categorizations in terms of principles and techniques, including classical schemes, remote schemes, and modern schemes~\cite{RN9}. All methods should adhere to a detection time of up to 2.0 seconds based on IEEE 1547-2018~\cite{ieee1547}.
Within classical methods, both passive and active techniques exist. In the passive method, changes in parameters using the measured signals are monitored\cite{reddy2020review}, whereas the active method injects power signal perturbations into the power grid to increase sensitivity~\cite{RN10}. Adaptive ROCOF islanding detection relay ~\cite{power1}, THD-based detection method~\cite{pd10}, impedance measurement based detection method~\cite{pd3}, An injection approach based on the maximum power point tracking concept is essential for islanding detection in photovoltaic systems. ~\cite{power2}, feasibility study of intentional islanding of conventional sources ~\cite{power3}, a statistical feature for islanding detection in the presence of inverters ~\cite{power4}, space vector domain islanding detection ~\cite{power5} are some of the methods available in this category.
In the second category, the remote schemes rely on information exchange while their processing technique is similar to classical methods~\cite{bekhradian2022comprehensive}. In the third category, modern methods analyze electrical signals using signal processing tools to identify islanding conditions~\cite{gupta2022review} and Artificial Intelligence (AI) based approaches utilize historical data to learn the islanding patterns~\cite{mohapatra2021comprehensive}. 

Recent works have emphasized the superiority of the AI-based approaches~\cite{RN11,pd8}. These methods can identify complex nonlinear systems and eliminate the need for constant threshold adjustments required by other techniques~\cite{RN2}. By using signal processing to extract features, intelligent computational methods can accurately detect critical scenarios, such as zero mismatches between active and reactive power, switching events, noisy systems, and grid faults. In the work proposed by Matic-Cuka et al.~\cite{matic2014islanding}, a technique using Support Vector Machines (SVM) is proposed for inverter-based distributed generation with 60 input features. Additionally, Vyas et al. ~\cite{vyas2016multivariate} employed a K-Nearest Neighbors (KNN) methodology for islanding detection in grid-tied solar PV systems. Such schemes have achieved an accuracy of up to 97\%. In recent years, recurrent neural networks (RNN) based on Long Short-Term Memory (LSTM) have been proposed\cite{massaoudi2021deep}, resulting in accuracy values exceeding 99\%. More precisely, Xia et al. ~\cite{xia2022novel} proposed a microgrid islanding detection using a multi-feature LSTM network, Abdelsalam et al.~\cite{9110832} included an inverter-based source, and in~\cite{(4)} a scheme based on harmonic signals within a multi-LSTM structure was used for islanding detection. More sophisticated approaches, e.g. in~\cite{(4)}, combined Convolutional neural networks (CNNs) and LSTM architectures to detect the islanding for a system with both inverter-based and synchronous-based sources. Despite achieving an islanding detection accuracy of over 99\%, these models suffer from a large computational complexity and instability in noisy conditions ~\cite{nois1,nois2,nois3,nois4,nois5,nois7,pd1}. For example, in~\cite{nois6,(2)} the islanding detection accuracy decreased to 97\% under Signal-to-Noise Ratio (SNR) conditions of 30 dB.

To address these shortcomings, this study introduces a novel method based on the Wavenet model for islanding detection \cite{vanwavenet}. WaveNet was originally designed by Google DeepMind for autoregressive audio generation through modeling the temporal dependencies within audio data. The backbone of our proposed model is WaveNet where we connect its last layer to a dense (fully connected) layer with a sigmoid activation function, thus enabling the prediction of the likelihood of islanding occurrence. Despite the simple structure of the WaveNet due to using 1D convolutional layers, the WaveNet classifier incorporates residual connection (or skip connection)  which makes it strong enough to mitigate the vanishing gradient problem, thus significantly enhancing the model's generalization capacity. The proposed scheme ensures high detection accuracy even in the presence of significant noise. This is achieved by integrating a feature denoising step, employing a U-Net model, into the detection algorithm. The comparative analysis demonstrates that our single Wavenet model exhibits greater robustness to noise, with a complexity reduced by more than half compared to existing state-of-the-art islanding detection models. Furthermore, the combination of Wavenet with U-Net denoising substantially enhances islanding detection accuracy, achieving up to 98\% accuracy under significant noise levels with a 10 dB signal-to-noise ratio (SNR). It is pertinent to underscore that the effectiveness of the suggested scheme will be rigorously evaluated through testing across various real-world scenarios, encompassing both islanding and non-islanding conditions. Moreover, the validation process will incorporate a realistic test system featuring conventional and inverter-based resources. The key contributions of this study are listed below:

\begin{enumerate}
    \item [-] We present a novel islanding detection model founded upon the deep generative Wavenet framework, wherein the final layer is adapted to facilitate islanding detection. Our findings demonstrate superior performance in comparison to state-of-the-art methodologies, coupled with a computational complexity reduced by more than half. To the authors' knowledge, This pioneering study introduces WaveNet for islanding classification for the first time.
    \item [-] We incorporate a feature denoising step, leveraging a pre-trained U-Net model, and demonstrate its substantial enhancement of the robustness of Wavenet-based islanding detection in environments characterized by severe noise. 
    \item [-] The performance of our Wavenet-based islanding detection, both standalone and in conjunction with U-Net feature denoising, is comprehensively evaluated against state-of-the-art methods across varying islanding and non-islanding scenarios at different signal-to-noise ratios (SNR). Our Wavenet model augmented with denoising U-Net exhibits a notable enhancement, achieving an accuracy exceeding 98\% at an SNR of 10 dB. 
\end{enumerate}

The subsequent sections of this paper are structured as follows: Sec. II outlines the test system and the specifics of the islanding simulation. Sec. III delves into the presentation of the deep islanding detection classifiers. Here, we introduce our Wavenet classifier alongside the U-Net denoising step aimed at enhancing islanding detection accuracy. The experimental findings of our proposed models are detailed in Sec. IV, where a comparative analysis with the LSTM islanding classifier is conducted, considering computational complexities and accuracy across varying noise levels. Finally, Sec. V offers concluding remarks to summarize the key findings and contributions of this study.

\emph{Notation and conventions.} We employ uppercase letters to denote random variables and lowercase letters to signify specific realizations or values. The notation $P(x)$ represents the probability distribution of the random variable $X$, while $\mathbb{E}[\cdot]$ denotes the expectation operator with respect to the joint distribution of all pertinent random variables. Additionally, $\mathbb{E}[Y|X]$ denotes the conditional expectation of $Y$ given $X$. For a sequence of random variables, or a time series, of length $T$, denoted by $X^T = (X_1, X_2, \dots, X_T)$, the corresponding realization is represented by $x^T = (x_1, x_2, \dots, x_T)$. The function $\text{inf}(\cdot)$ refers to the Infimum function, the sigmoid function is defined as $\sigma(z) = \frac{1}{1 + e^{-z}}$, and the hyperbolic tangent function is given by $\tanh(z) = \frac{1 - e^{-2z}}{1 + e^{-2z}}$.

%We use capital letters to denote random variables and lowercase to denote a realization or specific values. $P(x)$ is the probability distribution of random variable $X$, $\E[\cdot]$ is the expectation with respect to the joint distribution of all random variables involved, and $\E[Y|X]$ is the conditional expectation of $Y$ given $X$. $X^T=(X_1, X_2,\dots, X_T)$ is a sequence of random variables, or a time series, of length $T$ while $x^T=(x_1, x_2,\dots, x_T)$ is a realization of $X^T$. The inf(.) refers to the Infimum function, $\sigma (z) = \frac{1}{1+e^{-z}}$ is the sigmoid function, and $tanh(z) = \frac{1-e^{-2z}}{1+e^{-2z}}$ is the hyperbolic tangent function.

\section{Test System and Islanding Simulation} \label{sec:Test_system}

\begin{figure}[htbp]
	\centering
	\includegraphics[width=0.87\linewidth]{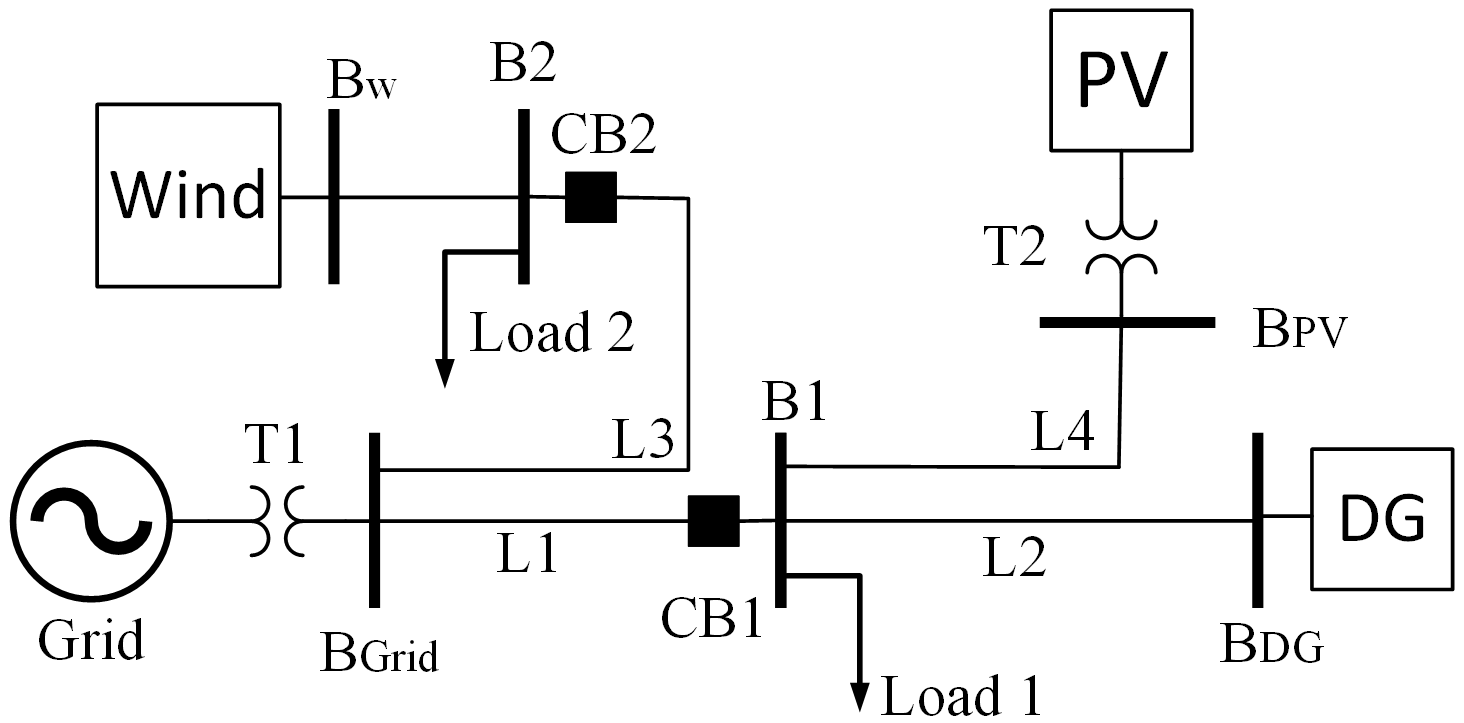}
	\caption{Schematic of the test system.}
	\label{tessy}
\end{figure}

The test network of this paper (which is similar to the one proposed in~\cite{nois6,myref}) comprises the Utility distribution system referred to as the Grid, a 250 kW Grid Connected Photovoltaic Plant denoted as the PV Unit, a 2 MW synchronous generator identified as DG, and a 1.5 MW DFIG Wind Turbine as Wind, all of which are detailed in Fig.~\ref{tessy} with the parameters listed in table~\ref{tab:test_m}.

The solar PV plant investigated in this study, with a capacity of 250 kW, is located 2 km away from $B_{1}$ Bus and linked to the PV Bus through a transformer. Its technical specifications are outlined in Table 1, comprising 86 parallel strings, each composed of seven modules of Sun Power SPR-415E with a series connection. The PV system employs a three-level IGBT leg with PWM unit control for conversion~\cite{n04}. Voltage transformation to 25 kVAC is accomplished by a 250 kVA 3$\phi$ 250V/25kV transformer. Additionally, another 250 kVA 25kV/400V 3$\phi$ transformer establishes the connection between the PV system and the PV Bus. The synchronous generator, a 3 MVA machine operating at 400 V, 50 Hz, and 1500 rpm, is driven by a diesel engine and linked to the DG bus, situated 500m away. Furthermore, a 2 MW DFIG Wind Turbine with variable speed operation is integrated with the Grid at $B_{w}$ Bus~\cite{n03}.

\begin{table}[htbp]
	\centering
	\caption{Parameters of the Test System}
	\begin{adjustbox}{width=0.47\textwidth}
		\begin{tabular}{c c}
			\toprule
            \textbf{Parameters}  & \textbf{Values} \\
            \midrule[0.1pt]
            
            PV arrays	& $86'times 7$ modules \\
            
            \midrule[0.1pt]

            PV modules&	SunPower SPR-415E\\
            
            \midrule[0.1pt]
            DG system	&3 MVA, 400 V, 50 Hz, 1500 rpm\\
            
            \midrule[0.1pt]
            T1	&25 kV/400 V; 10 MVA\\
            
            \midrule[0.1pt]
            T2	&25 kV/400 V; 5 MVA\\
            
            \midrule[0.1pt]
            L1,L3&	4 km\\
            
            \midrule[0.1pt]
            L4	&2 km\\
            
            \midrule[0.1pt]
            L2	&500 m\\
            
            \midrule[0.1pt]
            R1 and L1 for all lines &	1.273 m$\Omega$/km; 0.9337 mH/km\\
            
            \midrule[0.1pt]
            R0 and L0 for all lines &	38.64 m$\Omega$/km;  4.1264 mH/km\\
            
            \midrule[0.1pt]
            Load 1	&1.2 MW; 0.3 MVAR\\
            
            \midrule[0.1pt]
            Load 2&	0.6 MW; 0.2 MVAR\\

			\bottomrule
		\end{tabular}
		%\centering
	\end{adjustbox}
	\label{tab:test_m}
\end{table}

\subsection{Data gathering}

For the purpose of islanding detection, it is necessary to sample the electrical quantities of the PV bus including current and voltage samples, normally with the 1 kHz sampling frequency. Using the sampled data, some electrical quantities are extracted, which are used as supplementary features for detection processes. 
In this study, a set of six well-accepted electrical features was used, including positive/negative/zero sequences of voltages, rate of change of frequency (ROCOF), and superimposed voltage waveforms. For the selection of these features, the process began with an examination of over 60 primary features from previous studies, which included voltage, current, and power parameters, as well as their combinations and derivatives. Due to the complexity that would arise from using all features, a subset was chosen for testing and processing. To identify the most effective features, a combination of filter-based and wrapper-based methods was employed ~\cite{Fselection}. Initially, the features were filtered based on statistical characteristics, followed by an assessment of their performance. This approach streamlined the selection process, leading to the optimal feature combination for the best results. In the end, positive/negative/zero sequence components of voltages, superimposed voltage waveforms, and the rate of frequency change were selected for their superior performance and simplicity across various test scenarios.
%In this paper, a set of six well-accepted electrical features including positive/negative/zero sequences of voltages, rate of change of frequency (ROCOF), and superimposed voltage waveforms are used.

\begin{figure}[htbp]
	\centering
	\includegraphics[width=1\linewidth]{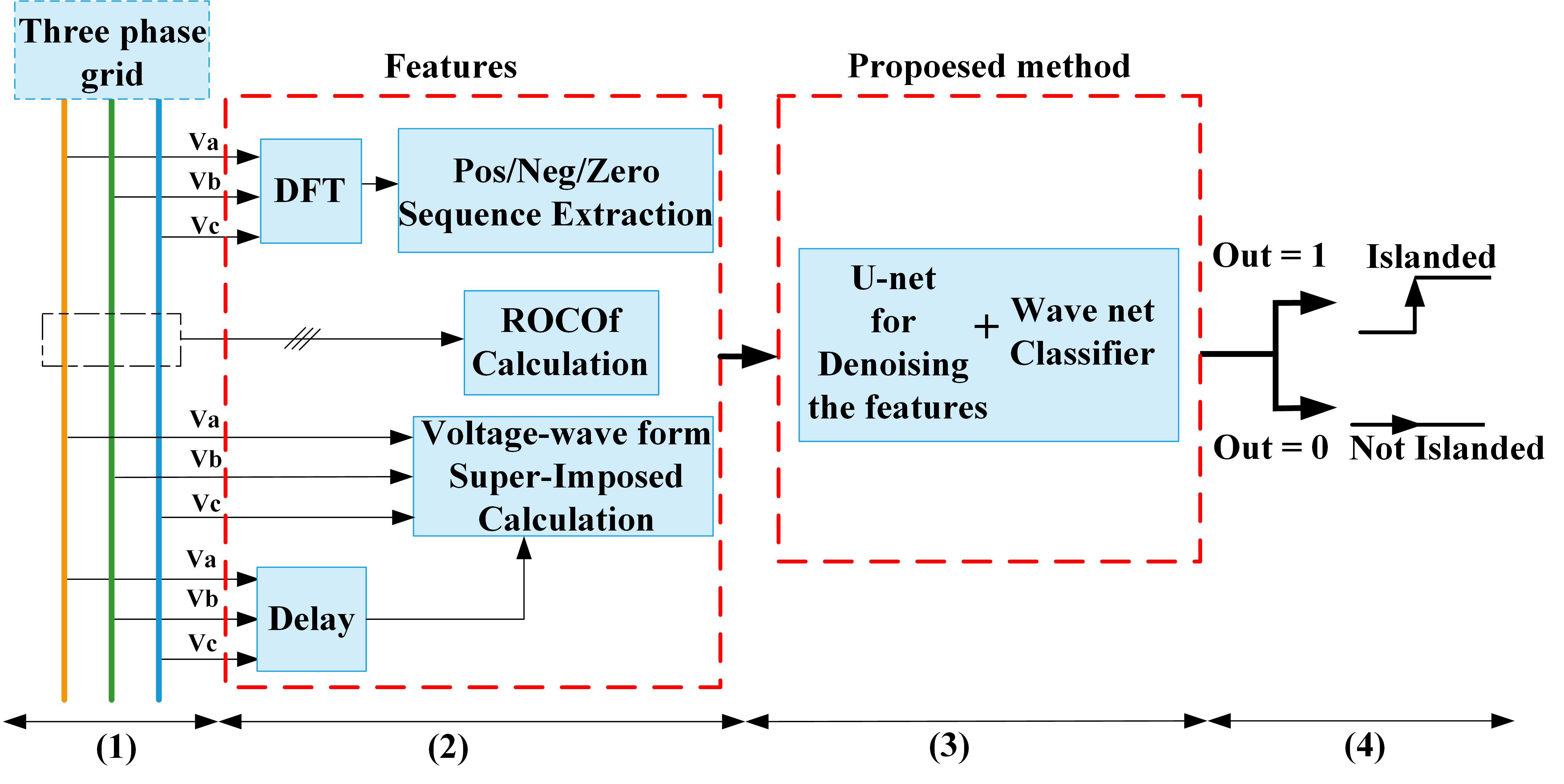}
	\caption{Block diagram of the proposed algorithm for islanding detection method: (1)- measuring three-phase voltages, (2)- feature calculation, (3)- Unet + WaveNet classifier, and (4)- islanding status signal output.}
	\label{flowchart}
\end{figure}

Fig.~\ref{flowchart} shows the schematic diagram of the suggested approach, where the measured three-phase voltage waveforms undergo processing by the feature extraction block to identify the necessary features for the proposed WaveNet+UNet scheme. Due to space constraints, the details of the feature extraction blocks are not extensively elaborated upon, as they are well-established in the current literature.

To assess the islanding scenarios and obtain the data, a test procedure is executed, which involves opening CB1 (grid-side circuit breaker) to isolate the grid from the main electrical network. This test is conducted across 50 distinct loads, each with varying active power levels, and 30 diverse reactive power settings. A total of 1500 load configurations were examined, all connected to the main bus $ \left (B_{1}\right )$. Also, non-islanding detection is evaluated by testing the subsequent scenarios, including switching a substantial linear load, connecting/disconnecting the capacitor banks, observing grid fault impacts, and modifying the Load Quality Factor $ \left ( L_{QF}\right )$. In this work, a total of 3080 time-series data samples are generated that include 2211 islanding cases and the rest non-islanding samples. Details regarding the data analysis and islanding detection will be discussed in section \ref{sec:results}.

\section{Proposed Deep Classifiers for Islanding Detection}\label{sec:wavenet}

In this section, the classification approach based on the deep neural networks for islanding detection is introduced. This network receives as input a time series which are features extracted from the voltage and current of an active distribution network and returns at the output the likelihood of islanding occurrence. More precisely, consider the time-series $X^T$ where for each $t\in\{1,2,\dots, T\}$, $X_t$ takes values on $\mathbb{R}^d$. Here $d$ refers to the number of features considered at time $t$. In addition, let's define the random variable $Y$, associated with each $X^T$, as the islanding label with $Y\in \mathbb{Y}=\{0,1\}$ where 1 refers to islanding. Thus, while $p(y|x^T)$ is the true probability distribution of the islanding labels given the time-series features $x^T$, the goal of the deep neural network classifier is to predict $q(y|x^T)$ as an approximation of the true distribution $p(y|x^T)$ through minimizing the Kullback-Leibler (KL) divergence~\cite{cover2006elements}:

\begin{align} \label{eq:optim}
\underset{q_{Y|X^T}}{\text{inf}} \;  \text{KL}\big (p_{Y X^T}\|q_{Y X^T}\big ) & = \underset{q_{Y|X^T}}{\text{inf}} \; \E\left[ \log \frac{p_{Y|X^T}(Y|X^T)}{q_{Y|X^T}(Y|X^T)} \right], 
\end{align}
where the expectation is taken concerning the true distribution $p_{Y X^T}$. Solving the optimization problem~\eqref{eq:optim} is equivalent to minimizing $\mathbb{E}[-\log q_{Y|X^T}(Y|X^T)]$, known as the cross-entropy. Throughout this study, the cross-entropy serves as the loss function, employed as the objective function.

%where the expectation is with respect to the true distribution $p_{Y X^T}$. Solving the optimization problem~\eqref{eq:optim} is equivalent to minimizing $\E[-\log q_{Y|X^T}(Y|X^T)]$, the so-called cross-entropy. Through this work, the cross-entropy is the loss function that will be used as the objective function.

\subsection{LSTM Classifier} 

\begin{figure}[htbp]
	\centering
	\includegraphics[width=1\linewidth]{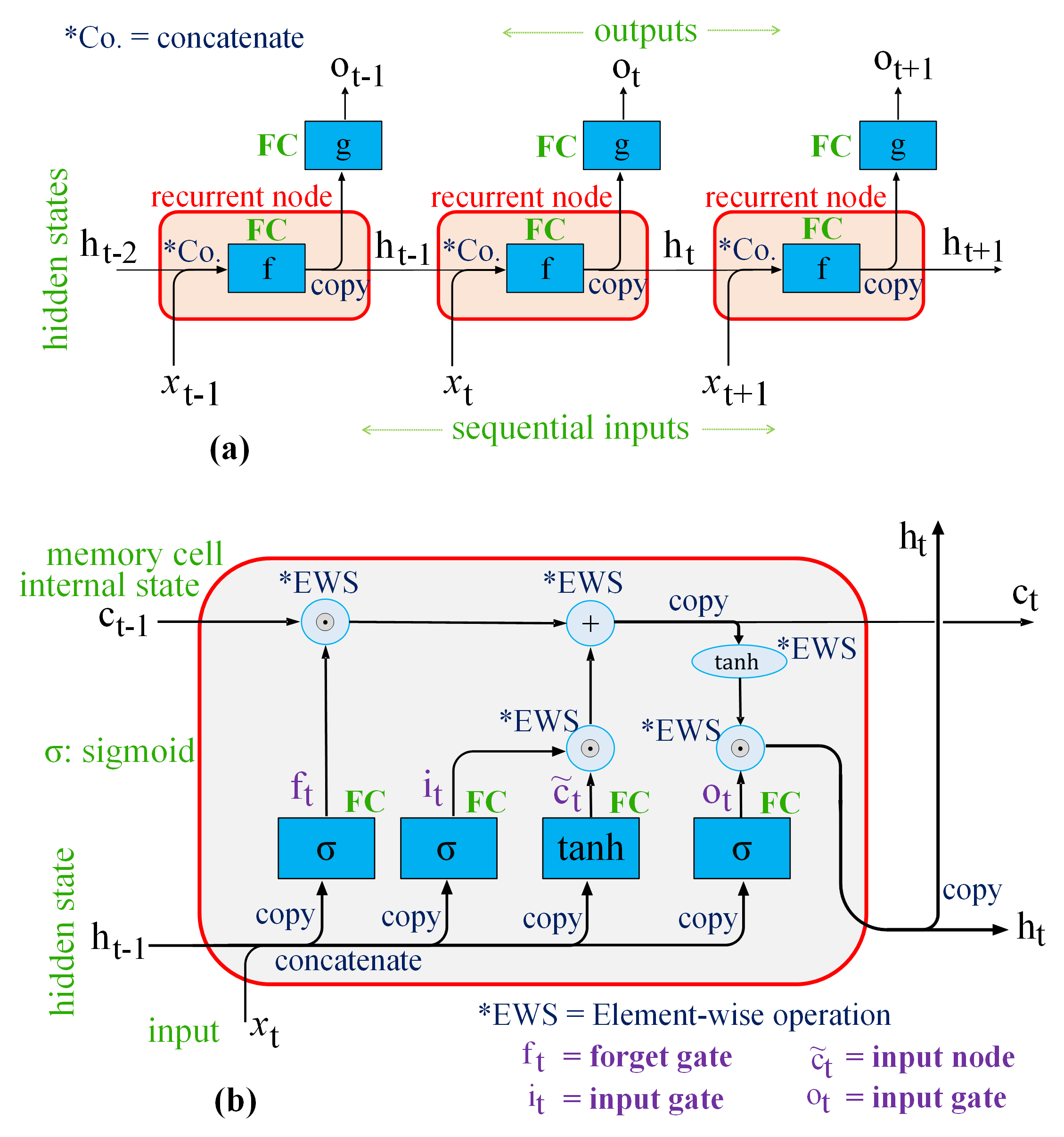}
	\caption{Schematic of (a) the recurrent neural network and (b) an LSTM cell diagram at time step t.}
	\label{lstm_rnn}
\end{figure}

Recurrent Neural Networks (RNNs) constitute a class of neural networks explicitly tailored for sequential data. Each RNN unit processes an input along with its hidden state from the previous time step thus enabling capturing the dynamics of sequential data over time. Fig.~\ref{lstm_rnn} (a) represents how the RNN works. In this figure, both networks $f$ and $g$ are modeled by fully connected (FC) layers. While RNNs are adept at capturing short-term dependencies, they struggle to retain information over longer sequences due to vanishing gradients~\cite{bengio1994learning}. This limits their practicality in scenarios requiring an understanding of context beyond recent history. Long Short-Term Memory (LSTM) Networks, introduced by Hochreiter and Schmidhuber in \cite{hochreiter1997long} and subsequently refined in \cite{gers1999learning}, enrich RNNs by integrating specialized memory cells and gating mechanisms. LSTMs are equipped with the capability to selectively store and retrieve information across extended sequences. (see Fig.~\ref{lstm_rnn} (b)). 

There are three key components that enable long-term modeling in LSTM including (1) the Forget gate that controls the memory's internal state and determines whether to keep it (important information) or flush it (remove the irrelevant information); (ii) the Output gate that determines the impact of the cell on the output at the current time; and (iii) Input gate and input node that determine how much information from the input should be added to the memory. After all, the output of the cell, i.e., the hidden state (which will be applied to the next time step) would be generated. The equations representing the formulations of LSTM nodes at time step t are as follows (equation~\eqref{LSTM_H}):

\begin{equation}\label{LSTM_H}
\begin{aligned}
    f_t &= \sigma(b^f + K^fh_{t-1} + V^f x_t)\\
    i_t &= \sigma(b^g + K^i h_{t-1} + V^i x_t)\\
    \Tilde{c}_t &= \sigma(b^{\Tilde{c}} + K^{\Tilde{c}} h_{t-1} + V^{\Tilde{c}} x_t)\\
    o_t &= \sigma(b^o + K^oh_{t-1} + V^ox_t)\\
    c_t &= f_t\times c_{t-1} + i_t\times i_t\\
    h_t &= o_t\times tanh\left(c_t\right)\\
\end{aligned}
\end{equation}

Here,  $b$, $K$, and $V$ represent biases, input weights, and recurrent weights associated with each gate, respectively. Stacking multiple LSTM layers and culminating in a fully connected layer with sigmoid activation significantly enhances the classification of the input time series.

%\noindent where the parameters $b, K, V$ are respectively biases, input weights, and recurrent weights associated with each gate. Stacking multiple LSTM layers ended with a fully connected layer with sigmoid activation enables classifying the input time-series.

\subsection{WaveNet Classifier}

\begin{figure}[htbp!]
	\centering
	\includegraphics[width=1\linewidth]{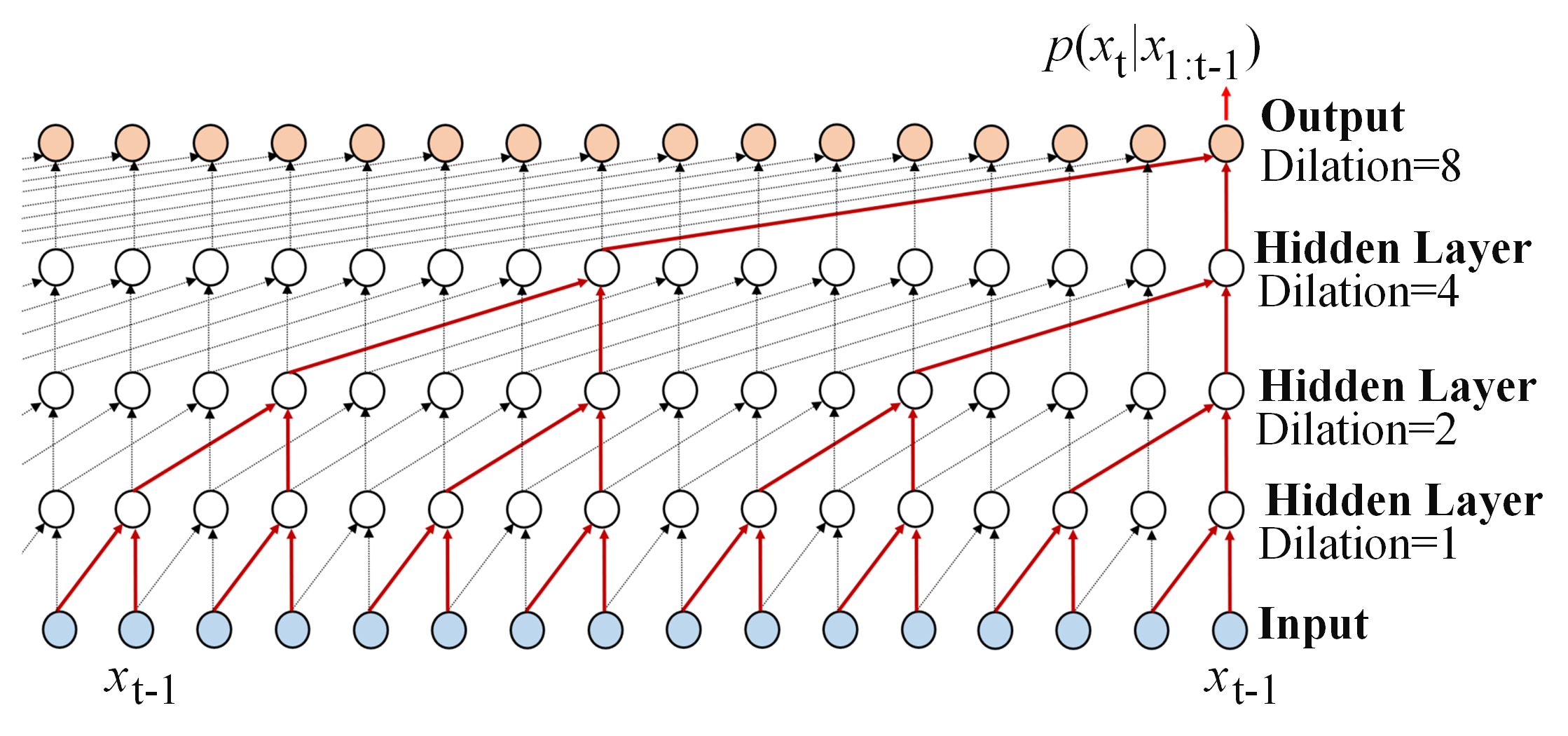}
	\caption{An example of the stack of causal convolutional layers inspired from~\cite{vanwavenet}. The depicted bold red arrows distinctly delineate the connections orchestrating the relationship between the output at time $t$ and the preceding layers. Notably, this configuration enables the acquisition of the conditional term $p(x_t|x_{1:t-1})$ based on the most recent 16 input time steps, corresponding to a receptive field of size 16.}
	\label{dilated}
\end{figure}

WaveNet is a generative model designed by DeepMind for synthesizing raw audio waveforms~\cite{vanwavenet}. Unlike traditional methods that generate audio by modeling spectrograms or other high-level features, WaveNet operates directly on the raw waveform data by autoregressive modeling of the audio probability distribution. More precisely, consider the sequence of random variables $X^T=(X_1, X_2,\dots, X_T)$ of length $T$ where for each $t\in\{1,2,\dots, T\}$, $X_t$ takes values on $\mathbb{R}^d$. Here $d$ refers to the number of features considered at time $t$. Using the chain rule of probability, $p(x^T)$ the probability distribution of sequential data $X^T$ can be written as follows:

\begin{equation} \label{eqtr1}
p(x^T) = p(x_1)p(x_2|x_1)\dots p(x_T|x_{1:T-1})= \prod_{t=1}^{T} p(x_t|x_{1:t-1})
\end{equation}

\noindent where we define $p(x_1|x_{1:0}) := p(x_1)$. The equation~\eqref{eqtr1} is called the autoregressive modeling of $p(x^T)$ and the WaveNet aims to learn $p(x^T)$ through modeling each of the conditional probability terms.  To this end, WaveNet employs a dilated causal convolutional architecture, inspired by the idea of dilated convolutions that effectively increase the receptive field (the number of previous times that impact the output at time $t$) without a corresponding increase in parameters. To understand the concept of dilated convolutions, consider a filter sliding over the input data, but instead of moving one step at a time, it jumps with increasing gaps. This is called a dilated convolution. These gaps, or dilation rates, determine how much context the model considers while making predictions. In addition, to ensure causality observed in equation~\eqref{eqtr1}, the WaveNet only uses information from the past to predict the future. Thus, the core of WaveNet is its stack of dilated causal convolutional layers (see Fig.~\ref{dilated}).

To improve the strength of the WaveNet model, gated activation units, an idea inspired by the LSTM gating mechanism is incorporated into designing WaveNet~\cite{vanwavenet}. More precisely, each convolutional layer is composed of a gated activation unit, which consists of two parallel branches – one for the filter and another for the gate. Assuming $x^T$ and $z^T$ as the input and output of the k$^{th}$ layer in the WaveNet network, the gated activation is modeled as follows:

\begin{equation} \label{eqtr2}
z^T = tanh(W_{f,k}\cdot x^T)\odot \sigma(W_{g,k}\cdot x^T)
\end{equation}

\noindent where $W_{f,k}$ and $W_{g,k}$ are the filter and gate parameters of the k$^{th}$ layer, and $\odot$ is the element-wise product or Hadamard product. The filter-branch learns to extract relevant features from the input, while the gate branch controls the information flow. This would help prevent vanishing gradients and allow the WaveNet model to capture complex dependencies. 

\begin{figure}[htbp]
	\centering
	\includegraphics[width=0.95\linewidth]{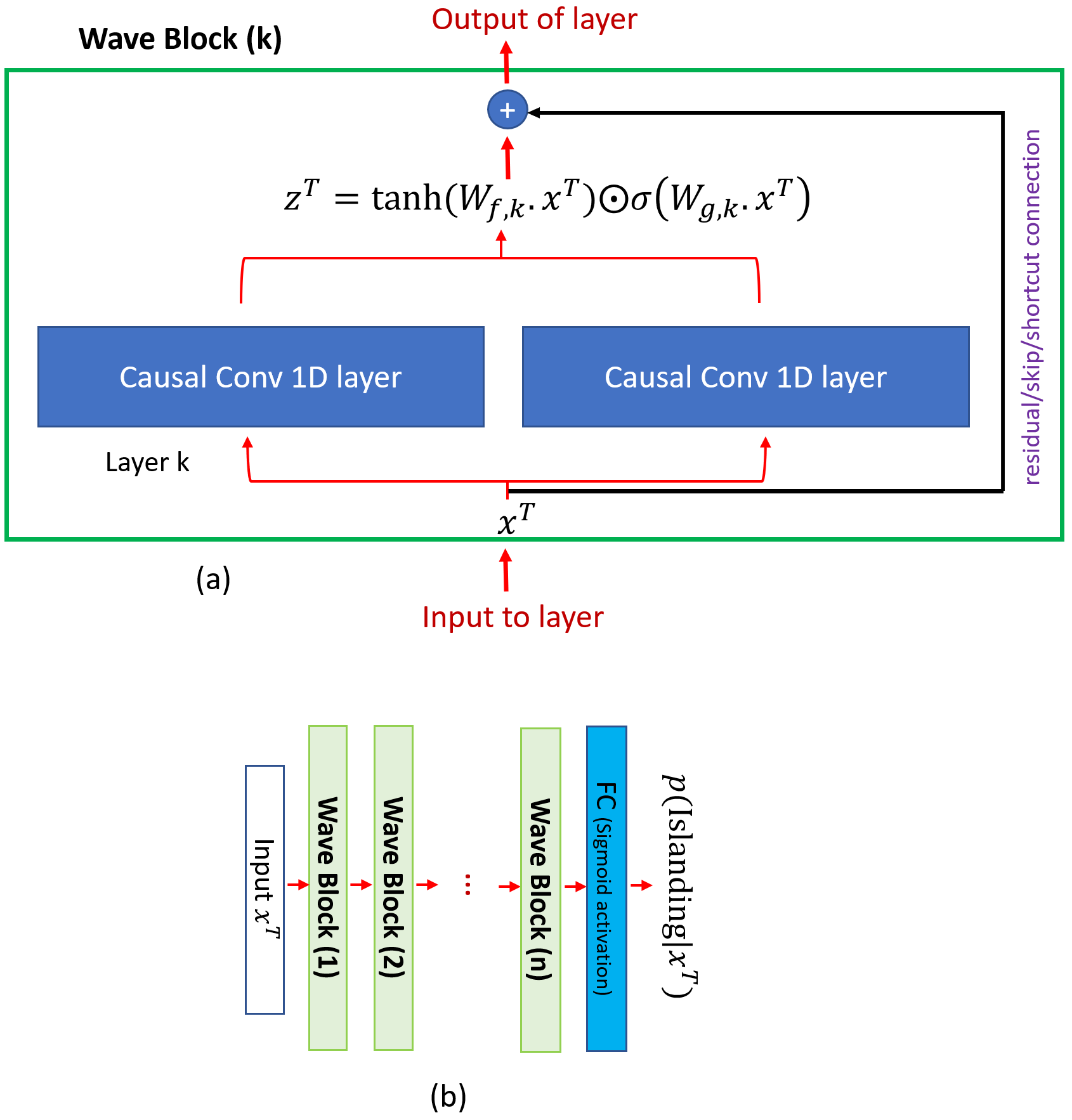}
	\caption{Schematic of (a) the gated activation units and residual connections used at each layer of the WaveNet, (b) the proposed WaveNet classifier for Islanding detection. In this model, the FC refers to a fully connected layer.}
	\label{wave_cl}
\end{figure}

On top of this gated activation, to facilitate the learning of residual representations while simultaneously enhancing the ability to capture fine-grained features, skip connections, also known as residual connections, are utilized~\cite{he2016deep}. This simple idea enables the direct transfer of information from input to output and thus speeds up the convergence while enabling the training of a deeper WaveNet model (see Fig.~\ref{wave_cl} (a)).

So far, the WaveNet for generating time-series samples (e.g. audio) was presented. To enable classification, the output of the WaveNet model is connected to a fully connected layer with one output and a Sigmoid activation function. The whole model is presented in Fig.~\ref{wave_cl} (b).

\subsection{Feature Denoising with U-NET Model}
\begin{figure}[htbp]
	\centering
	\includegraphics[width=0.99\linewidth]{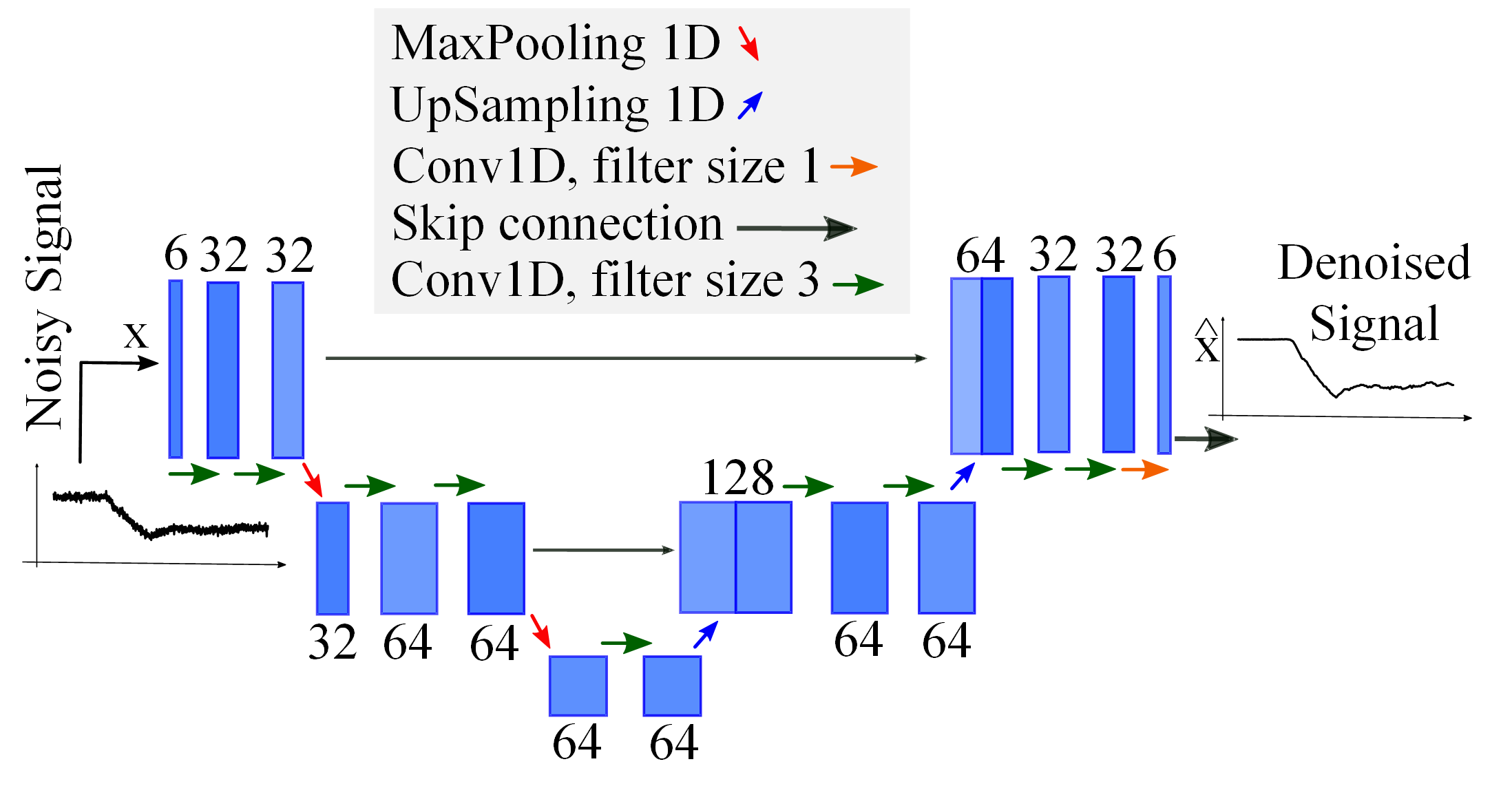}
	\caption{U-NET framework for denoising islanding features. The numbers at the output of each layer represent the size of the feature map. Note that in our pre-trained U-Net model, the length of the input features is reduced by a factor of two at each U-Net level in the encoding path and then it is increased by a factor of two at each level of the decoding path.}
	\label{U-Net}
\end{figure}

The U-Net architecture, initially proposed by Ronneberger et al. \cite{ronneberger2015u} for biomedical image segmentation, has been adapted for various image processing tasks, including signal denoising. This paper presents an adaptation of the U-Net model tailored for 1D signal denoising using convolutional neural networks (CNNs). Comprising a contracting path (or encoder) followed by an expansive path (or decoder), the U-Net architecture efficiently captures context and reduces spatial dimensions through the contracting path, while enabling precise localization via the expansive path (see Fig. \ref{U-Net}).

The contracting path comprises multiple layers of convolutional and max-pooling operations. Here, $x$ denotes the input signal, and $C_k$ represents the $k$-th convolutional layer with a kernel size of $3 \times 3$ and a rectified linear unit (ReLU) activation function denoted by $\sigma$:

\begin{equation}
  C_k(x) = \sigma(W_k \cdot x + b_k)  
\end{equation}

\noindent where $W_k$ and $b_k$ are the learnable weights and biases of the $k$-th convolutional layer, respectively. The max-pooling operation reduces the spatial dimensions of the feature maps.

The expansive path consists of upsampling followed by convolutional operations. Let $U_k$ represent the $k$-th upsampling layer, and $D_k$ denote the $k$-th convolutional layers are followed by upsampling layers, which definitively increase the spatial resolution of the feature maps.

\begin{equation}
\begin{aligned}
    U_k(x) &= \text{Upsampling}(x) \\
     D_k(x) &= \sigma(W'_k \cdot x + b'_k)   
\end{aligned}
\end{equation}

\noindent where $W'_k$ and $b'_k$ are the learnable weights and biases of the $k$-th convolutional layer in the expansive path, respectively.

For precise localization, skip connections are essential as they concatenate feature maps from the contracting path with corresponding feature maps in the expansive path. 

Finally, the U-NET model is trained using stochastic gradient descent (SGD) algorithms to minimize the mean absolute error (or mean squared error) loss function.

\section{Results and Discussion} \label{sec:results}

\subsection{Data Simulation and Models Structures}

This study employs a dataset of a total of 3080 time-series data samples, each spanning a length of 10 ms and encompassing six features, as detailed in Section II. The dataset is partitioned into two distinct sets: training and test datasets, with an 80-20 ratio. Throughout the training phase, $20\%$ of the training data samples serve as validation data to fine-tune the hyperparameters associated with each model, such as learning rate, number of layers, and number of nodes in each layer. The details regarding the models' structures are listed in table~\ref{tab:result_total_metrics}.

\begin{table*}[htbp]
	\centering
	\caption{Comprehensive comparison of the performance of various models based on evaluation metrics applied to test data samples. For each model, the average and standard deviation of five runs are presented.}
	\begin{adjustbox}{width=0.99\textwidth}
		\begin{tabular}{l c c c c c}
			\toprule
            \textbf{Model}  & \textbf{Model complexity} & \textbf{F1-score} &\textbf{Balanced accuracy}&\textbf{Precision}& \textbf{Recall}\\
            \midrule[0.1pt]
            
            WaveNet classifier (ours)&\makecell{Total of five Wave Blocks where each\\ has 2 convolutional layers with filter size 32\\ (Training parameters: 63,105)\\ Adam optimizer with learning rate $0.0002$} &$0.9989\pm0.0007$&$0.9981\pm0.0015$&$0.9991\pm0.0011$&$0.9987\pm0.0011$\\
            
            \midrule[0.1pt]
            
            LSTM classifier (state-of-the-art~\cite{9110832, nois6})& \makecell{4 LSTM layers with 64, 128, 64, and 32 cells\\(Training parameters: 178,849)\\ RMSprop optimizer with learning rate $0.001$}&$0.9915\pm0.0057$&$0.9863\pm0.0046$&$0.9938\pm0.0032$&$0.9893\pm0.0133$\\

			\bottomrule
		\end{tabular}
		%\centering
	\end{adjustbox}
	\label{tab:result_total_metrics}
\end{table*}

In the context of our investigation, we categorize the islanding condition as the positive class (Class 1), while the non-islanding case serves as the negative class (Class 0). This binary classification setup allows us to analyze two distinct types of errors that each classifier might make. The first error type corresponds to a false positive (FP), signifying a false indication of islanding. Conversely, the second error type represents a false negative (FN), denoting a failure to detect an actual islanding event. Assessing the effectiveness of any classifier hinges upon minimizing these two error types. Commonly, this optimization objective is achieved by maximizing accuracy, as defined below:

\begin{equation*}\label{eq:acc}
    \text{accuracy} = \frac{TP+TN}{TP + TN + FP + FN}
\end{equation*}

\noindent where TP signifies true positives, and TN denotes true negatives. However, in scenarios characterized by imbalanced datasets (e.g. in our case where the test dataset includes 448 islanding and 168 non-islanding), accuracy can be misleading. The reason lies in the potential for a no-skill model, which arbitrarily assigns the majority class to all instances, to exhibit high accuracy. To address this issue and provide a more comprehensive evaluation, alternative metrics are used~\cite{saito2015precision}:

\begin{align}\label{eq:metrics}
    &\text{Balanced accuracy} = 0.5\times\left[\frac{TP}{TP+FN} + \frac{TN}{TN+FP}\right] \nonumber\\
    &\text{Precision} = \frac{TP}{TP + FP} \nonumber\\
    &\text{Recall} = \frac{TP}{TP + FN} \\
    &\text{F1-score} = 2\times\frac{\text{Precision}\times \text{Recall}}{\text{Precision}+ \text{Recall}}\nonumber
\end{align}

To rigorously evaluate the performance of the proposed models in our study, we will employ the metrics outlined in equation~\eqref{eq:metrics}. These metrics provide a much clearer understanding of the classifiers' capabilities in the presence of imbalanced datasets and varying error costs.

\subsection{WaveNet Islanding Detection}

\begin{figure}[htbp!]
    \centering
    \includegraphics[width=0.6\linewidth]{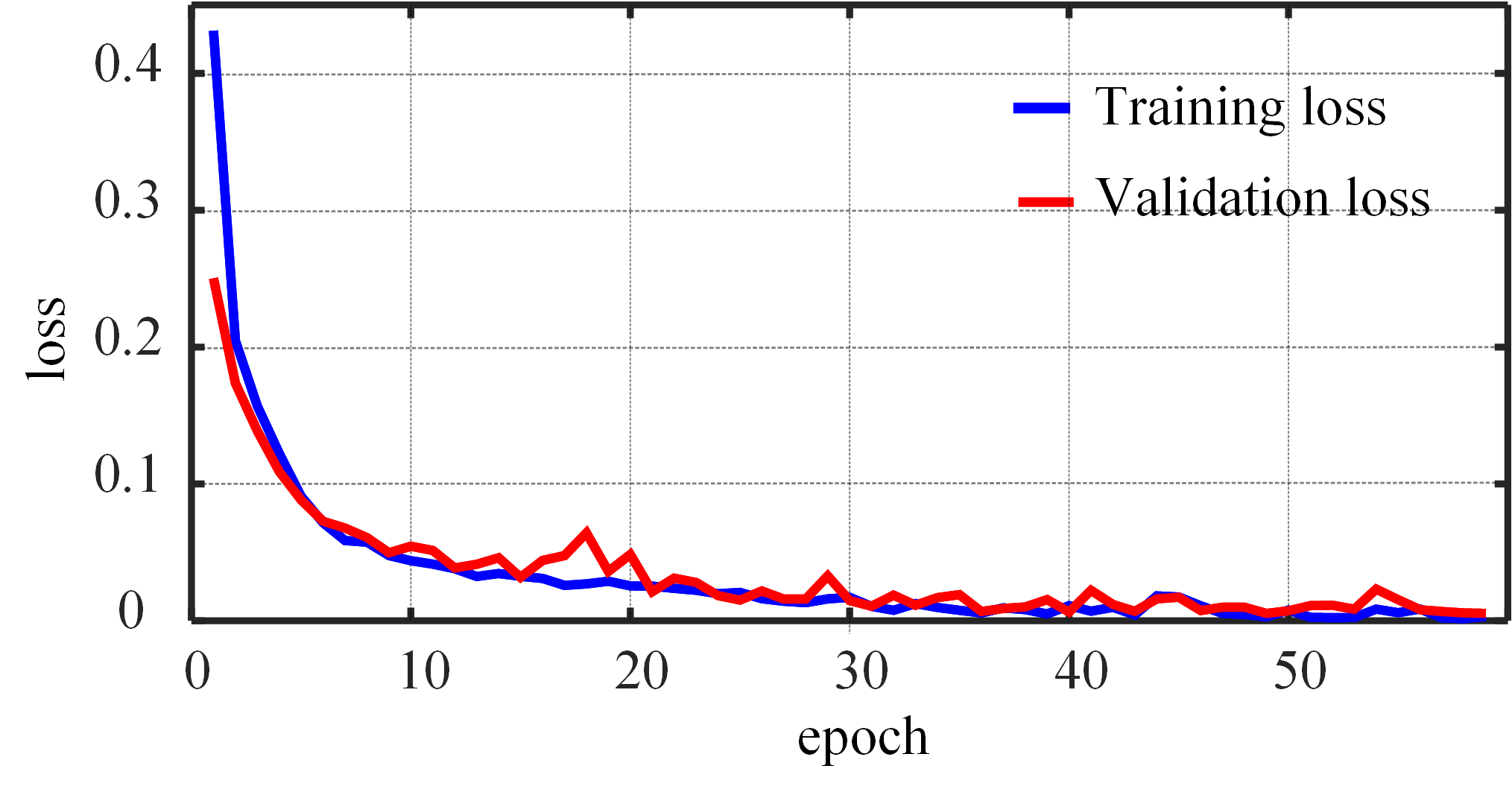}
    \caption{An exemplar representation showcasing the loss of model during the training of the WaveNet classifier.}
    \label{fig:loss}
\end{figure}

In assessing the effectiveness of the WaveNet classifier model, our foremost consideration is to prevent overfitting. To accomplish this, we meticulously monitor the model's performance. To this end, Figure~\ref{fig:loss} visually captures the evolution of the training and validation loss functions across training epochs. Furthermore, as a precautionary measure, we implement early stopping during training with a patience parameter set to 10. This ensures that the model's learning process is terminated If no improvement is seen over a specified number of epochs, effectively safeguarding against the potential pitfalls of overfitting. Analysis of Fig.~\ref{fig:loss} reveals that the WaveNet Model exhibits resilience against overfitting. Evidently, the loss functions demonstrate a convergence to stability after a certain number of training epochs. This observation underscores the model's capacity to generalize effectively, thus contributing to its suitability for the given task. 

The comprehensive outcomes of all models, as gauged by the evaluation metrics, are concisely presented in Table~\ref{tab:result_total_metrics}. These results are confidently based on the mean values and one standard deviation derived from five distinct runs. Notably, an insightful observation emerges from this table: the novel WaveNet classifier model, which boasts a significantly lower model complexity ( i.e. $63,105$ learning parameters for the WaveNet model compared with $178,849$ learning parameters for the LSTM model), notably surpasses the state-of-the-art LSTM classifier in performance. In addition, it can be seen that the standard deviation of the evaluation measures for the LSTM classifier is much higher than the WaveNet classifier. It means the WaveNet is much more stable and has a better generalization.

\begin{figure}[htbp!]
    \centering
    \includegraphics[width=1\linewidth]{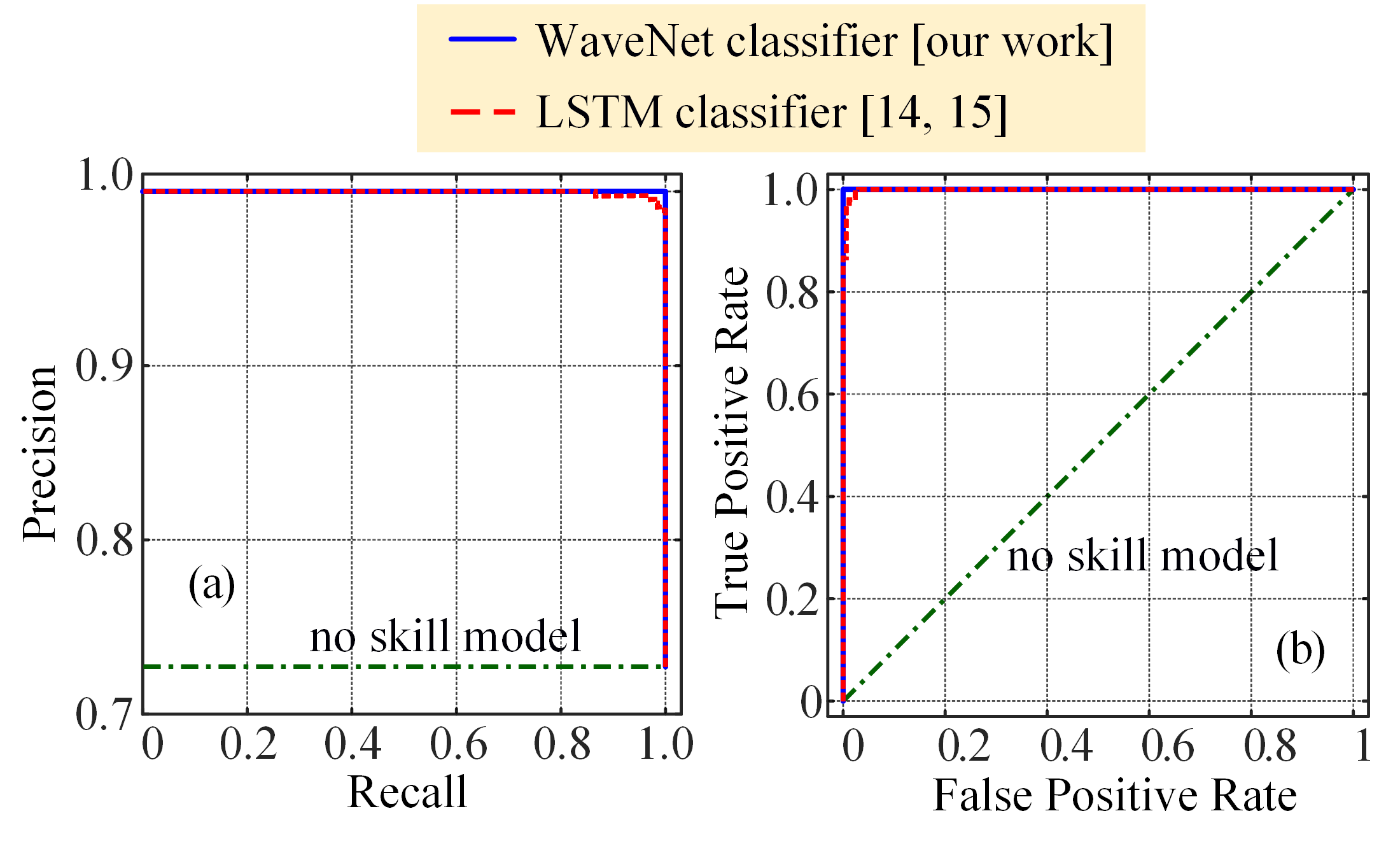}
    \caption{A comparison between the proposed WaveNet classifier and LSTM classifier \cite{9110832, xia2022novel} conducted on test data samples, focusing on two curves: (a) the precision-recall curve and (b) the receiver operating characteristic curve. To establish a baseline, we are considering the no-skill model, which is a dummy classifier that randomly assigns class labels without considering the input. As a result, the no-skill model has a true positive rate that is equal to its false positive rate. Its precision-recall curve appears as a horizontal line, representing the proportion of positive (islanding) samples compared to the total.}
    \label{fig:prec_roc}
\end{figure}

To further illustrate the contrasting performances of these models, we turn to the visual representations of their Receiver Operating Characteristics (ROC) curves and Precision-Recall curves, depicted in Fig.~\ref{fig:prec_roc}. More specifically, Fig.~\ref{fig:prec_roc} (a) represents that the WaveNet classifier has the highest area under the curve (AUC) while from Fig.~\ref{fig:prec_roc} (b) it can be seen that the WaveNet classifier not only has the higher AUC but also has a higher TPR for small FPR values.

\subsection{Islanding Detection with Features Denoising U-NET}

In this section, the robustness of the proposed models is investigated empirically against noisy features. To this end, the WaveNet and LSTM classifiers are tested with noisy features to determine islanding cases. Table~\ref{tab:result_noise} lists the performances of these models at different levels of noise based on balanced accuracy. As expected, noisy features, especially after SNR 15dB, degrade extensively the models' performance although the WaveNet classifier looks more robust to noisy features compared to the LSTM classifier. 

In response to this, we equip the WaveNet with a pre-trained U-NET denoising. In this work, a U-NET model consisting of an encoder-decoder framework with skip connections to preserve spatial information during downsampling and upsampling operations is used.
In the encoder part, two consecutive 1D convolutional layers with 32 filters each and ReLU activation function are employed, followed by max-pooling with a pool size of 2 to downsample the input signals effectively. 
After that, two additional convolutional layers with 64 filters each and ReLU activation are applied, followed by max-pooling.
This stage further abstracts features and serves as the bottleneck of the network. The decoder confidently upsamples the feature maps to the original input resolution.
It starts with upsampling followed by concatenation with feature maps from the corresponding encoder stage. Then, two convolutional layers with 64 filters each and ReLU activation are applied. Another upsampling operation is performed, followed by concatenation and two more convolutional layers with 32 filters each and ReLU activation. Finally, a 1D convolutional layer with 6 filters (number of features) generates the denoised output signals. In this work, the U-NET model is pre-trained at SNR 15dB and is used for all the levels of noise where the Adam optimizer with a learning rate of 0.0002 optimizes the U-NET parameters by minimizing the mean absolute error (MAE) loss function.

\begin{figure}[htbp!]
    \centering
    \includegraphics[width=0.6\linewidth]{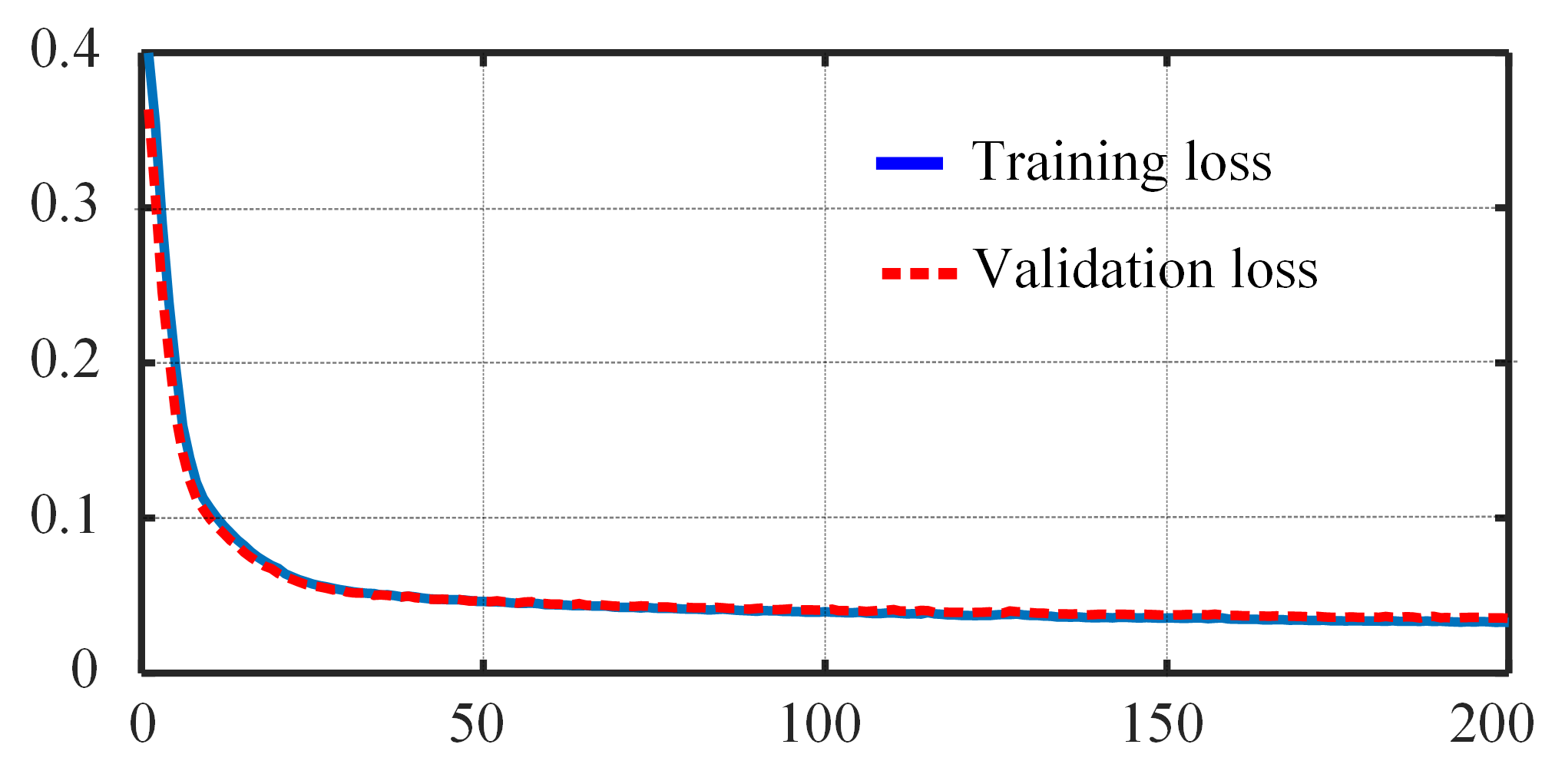}
    \caption{The loss function of U-NET denoising model versus training epochs. The tinny gap between the training and validation losses ensures that the model is not overfitting.}
    \label{fig:U-Netloss}
\end{figure}

In figure \ref{fig:U-Netloss}, it is clear that the U-NET model can minimize the MAE loss function without overfitting. In addition, the results of islanding detection with U-NET feature denoising at different levels of the noise are presented in Table~\ref{tab:result_noise}. From this table, it can be seen that our proposed $\text{WaveNet+U-NET}$ model can significantly improve the islanding detection accuracy even when the features are extensively noisy.

\begin{table*}[htbp]
	\centering
	\caption{Comparing models based on balanced accuracy in the presence of noise. Results show performances on test data samples. Average and one standard deviation of five runs are displayed for each model.}
	\begin{adjustbox}{width=0.85\textwidth}
		\begin{tabular}{l c c c c}
			\toprule
            \textbf{Model}  & \textbf{20 dB} &\textbf{15 dB}&\textbf{10 dB}& \textbf{5 dB}\\
            \midrule[0.1pt]
            
            LSTM classifier (state-of-the-art~\cite{9110832, nois6})&$0.9746 \pm 0.0031$  &$0.9630 \pm 0.0056$ &$0.9377 \pm 0.0110$ & $0.9008 \pm 0.0148$\\
            
            \midrule[0.1pt]

            WaveNet classifier (ours)&$0.9892 \pm 0.0031$ &$0.9729 \pm 0.0074$  &$0.9388 \pm 0.0101$ &$0.9102 \pm 0.0112$ \\

            \midrule[0.1pt]
            
            WaveNet classifier with denoising U-NET (ours)& $0.9927 \pm 0.0047$ &$0.9903 \pm 0.0075$ &$0.9837 \pm 0.0076$ &$0.9540 \pm 0.0093$ \\

			\bottomrule
		\end{tabular}
		%\centering
	\end{adjustbox}
	\label{tab:result_noise}
\end{table*}

\section{Conclusion} \label{sec:conclusion}
In this paper, an innovative WaveNet classifier equipped with a pre-trained denoising U-Net was introduced for islanding detection in active distribution networks. Unlike state-of-the-art e.g. LSTM model, which may encounter challenges like gradient vanishing and instability, the WaveNet classifier employs simple 1D convolutional structure layers with skip connections to enhance prediction accuracy while ensuring stability. This approach reduces parameter usage compared to LSTM, thereby increasing the efficiency and simplicity of the detection system and ultimately lowering implementation costs. To tackle the issue of islanding detection in noisy grids, a U-Net model with an encoder-decoder framework and skip connections is utilized. This strategy guarantees the robustness of the detection algorithm against significant system noise, achieving an unprecedented accuracy of 98.37\% even under 10 dB noisy conditions. The validation of this method involved testing various real-world scenarios representing both islanding and non-islanding conditions, encompassing different load active/reactive power values, load switching transients, capacitor bank switching, fault conditions in the main grid, diverse load quality factors, and varying signal-to-noise levels. The incorporation of a realistic test system and the integration of both conventional and inverter-based resources further bolster the credibility of the analysis. Overall, our proposed model showcases a substantial improvement in islanding detection accuracy, particularly in noisy environments. Finally, a comparison is drawn with the state-of-the-art LSTM classifier, demonstrating the superiority of the proposed scheme in terms of lower model complexity (i.e., 63,105 learning parameters for WaveNet compared to 178,849 learning parameters for LSTM) as well as stability and generalization (significantly lower standard deviation of evaluation measures compared to LSTM).

\bibliographystyle{ieeetr}
\bibliography{HREF}

\begin{thebibliography}{10}

\bibitem{RN1}
B.~K. Chaitanya, A.~Yadav, M.~Pazoki, and A.~Y. Abdelaziz, {\em Chapter 8 - A comprehensive review of islanding detection methods}, pp.~211--256.
\newblock Academic Press, 2021.

\bibitem{n01}
S.~Dutta, P.~K. Sadhu, M.~J.~B. Reddy, and D.~K. Mohanta, ``Shifting of research trends in islanding detection method - a comprehensive survey,'' {\em Protection and Control of Modern Power Systems}, vol.~3, no.~1, pp.~1--20, 2018.

\bibitem{mumtaz2023extensive}
F.~Mumtaz, K.~Imran, A.~Abusorrah, and S.~B.~A. Bukhari, ``An extensive overview of islanding detection strategies of active distributed generations in sustainable microgrids,'' {\em Sustainability}, vol.~15, no.~5, p.~4456, 2023.

\bibitem{pd4}
L.~Chen, X.~Dong, B.~Wang, L.~Shang, and C.~Liu, ``An edge computing-oriented islanding detection using differential entropy and multi-support vector machines,'' {\em IEEE Transactions on Smart Grid}, 2023.

\bibitem{n02}
O.~Dharmapandit, R.~K. Patnaik, and P.~K. Dash, ``A fast time-frequency response based differential spectral energy protection of ac microgrids including fault location,'' {\em Protection and Control of Modern Power Systems}, vol.~2, no.~4, pp.~1--28, 2017.

\bibitem{RN9}
O.~Arguence, F.~Cadoux, B.~Raison, and L.~D. Alvaro, ``Impact of power regulations on unwanted islanding detection,'' {\em IEEE Transactions on Power Electronics}, vol.~33, no.~10, pp.~8972--8981, 2018.

\bibitem{ieee1547}
D.~G. Photovoltaics and E.~Storage, ``Ieee standard for interconnection and interoperability of distributed energy resources with associated electric power systems interfaces,'' {\em IEEE std}, vol.~1547, pp.~1547--2018, 2018.

\bibitem{reddy2020review}
C.~R. Reddy, B.~S. Goud, B.~N. Reddy, M.~Pratyusha, C.~V. Kumar, and R.~Rekha, ``Review of islanding detection parameters in smart grids,'' in {\em 2020 8th International Conference on Smart Grid (icSmartGrid)}, pp.~78--89, IEEE, 2020.

\bibitem{RN10}
R.~Bakhshi-Jafarabadi, J.~Sadeh, A.~Serrano-Fontova, and E.~Rakhshani, ``Review on islanding detection methods for grid-connected photovoltaic systems, existing limitations and future insights,'' {\em IET Renewable Power Generation}, vol.~16, no.~15, pp.~3406--3421, 2022.

\bibitem{power1}
M.~Grebla, J.~R. Yellajosula, and H.~K. H{\o}idalen, ``Adaptive frequency estimation method for rocof islanding detection relay,'' {\em IEEE Transactions on Power Delivery}, vol.~35, no.~4, pp.~1867--1875, 2019.

\bibitem{pd10}
R.~Chandrakar, R.~K. Dubey, and B.~K. Panigrahi, ``Thd-based passive islanding detection technique for droop-controlled grid-forming inverters,'' {\em IEEE Systems Journal}, 2023.

\bibitem{pd3}
J.~Xing and L.~Mu, ``A novel islanding detection method for distributed pv system based on $\mu$pmus,'' {\em IEEE Transactions on Smart Grid}, 2023.

\bibitem{power2}
R.~Bakhshi-Jafarabadi, J.~Sadeh, and M.~Popov, ``Maximum power point tracking injection method for islanding detection of grid-connected photovoltaic systems in microgrid,'' {\em IEEE Transactions on Power Delivery}, vol.~36, no.~1, pp.~168--179, 2020.

\bibitem{power3}
R.~L. Lima, J.~P. Bonaldo, J.~C. Vieira, and R.~M. Monaro, ``A graphical method to assess the technical feasibility of intentional islanding of distributed synchronous generators,'' {\em IEEE Transactions on Power Delivery}, vol.~38, no.~1, pp.~742--745, 2022.

\bibitem{power4}
Y.~Bansal and R.~Sodhi, ``A statistical features based generic passive islanding detection scheme for iidgs system,'' {\em IEEE Transactions on Power Delivery}, vol.~37, no.~4, pp.~3176--3188, 2021.

\bibitem{power5}
M.~R. Alam, M.~T.~A. Begum, and B.~Mather, ``Islanding detection of distributed generation using electrical variables in space vector domain,'' {\em IEEE Transactions on Power Delivery}, vol.~35, no.~2, pp.~861--870, 2019.

\bibitem{bekhradian2022comprehensive}
R.~Bekhradian and M.~Sanaye-Pasand, ``A comprehensive survey on islanding detection methods of synchronous generator-based microgrids: Issues, solutions and future works,'' {\em IEEE Access}, vol.~10, pp.~76202--76219, 2022.

\bibitem{gupta2022review}
N.~Gupta, R.~Dogra, R.~Garg, and P.~Kumar, ``Review of islanding detection schemes for utility interactive solar photovoltaic systems,'' {\em International Journal of Green Energy}, vol.~19, no.~3, pp.~242--253, 2022.

\bibitem{mohapatra2021comprehensive}
S.~S. Mohapatra, M.~K. Maharana, and S.~B. Pati, ``Comprehensive review to analyze the islanding in distributed generation system,'' in {\em 2021 1st International Conference on Power Electronics and Energy (ICPEE)}, pp.~1--7, IEEE, 2021.

\bibitem{RN11}
A.~Hussain, C.~H. Kim, and A.~Mehdi, ``A comprehensive review of intelligent islanding schemes and feature selection techniques for distributed generation system,'' {\em IEEE Access}, vol.~9, pp.~146603--146624, 2021.

\bibitem{pd8}
M.~Karimi, M.~Farshad, R.~Azizipanah-Abarghooee, and K.~Kauhaniemi, ``Harmonic signature-based one-class classifier for islanding detection in microgrids,'' {\em IEEE Systems Journal}, 2023.

\bibitem{RN2}
M.-S. Kim, R.~Haider, G.-J. Cho, C.-H. Kim, C.-Y. Won, and J.-S. Chai, ``Comprehensive review of islanding detection methods for distributed generation systems,'' {\em Energies}, vol.~12, no.~5, 2019.

\bibitem{matic2014islanding}
B.~Matic-Cuka and M.~Kezunovic, ``Islanding detection for inverter-based distributed generation using support vector machine method,'' {\em IEEE Transactions on Smart Grid}, vol.~5, no.~6, pp.~2676--2686, 2014.

\bibitem{vyas2016multivariate}
S.~Vyas, R.~Kumar, R.~Kavasseri, {\em et~al.}, ``Multivariate statistics and supervised learning for predictive detection of unintentional islanding in grid-tied solar pv systems,'' {\em Applied Computational Intelligence and Soft Computing}, vol.~2016, 2016.

\bibitem{massaoudi2021deep}
M.~Massaoudi, H.~Abu-Rub, S.~S. Refaat, I.~Chihi, and F.~S. Oueslati, ``Deep learning in smart grid technology: A review of recent advancements and future prospects,'' {\em IEEE Access}, vol.~9, pp.~54558--54578, 2021.

\bibitem{xia2022novel}
Y.~Xia, F.~Yu, X.~Xiong, Q.~Huang, and Q.~Zhou, ``A novel microgrid islanding detection algorithm based on a multi-feature improved lstm,'' {\em Energies}, vol.~15, no.~8, p.~2810, 2022.

\bibitem{9110832}
A.~A. Abdelsalam, A.~A. Salem, E.~S. Oda, and A.~A. Eldesouky, ``Islanding detection of microgrid incorporating inverter based dgs using long short-term memory network,'' {\em IEEE Access}, vol.~8, pp.~106471--106486, 2020.

\bibitem{(4)}
A.~K. {\"O}zcanl{\i} and M.~Baysal, ``A novel multi-lstm based deep learning method for islanding detection in the microgrid,'' {\em Electric Power Systems Research}, vol.~202, p.~107574, 2022.

\bibitem{nois1}
N.~K. Swarnkar, O.~P. Mahela, B.~Khan, and M.~Lalwani, ``Identification of islanding events in utility grid with renewable energy penetration using current based passive method,'' {\em IEEE Access}, vol.~9, pp.~93781--93794, 2021.

\bibitem{nois2}
A.~Y{\i}lmaz and G.~Bayrak, ``A new signal processing-based islanding detection method using pyramidal algorithm with undecimated wavelet transform for distributed generators of hydrogen energy,'' {\em International Journal of Hydrogen Energy}, vol.~47, no.~45, pp.~19821--19836, 2022.

\bibitem{nois3}
S.~Chandak, M.~Mishra, S.~Nayak, and P.~K. Rout, ``Optimal feature selection for islanding detection in distributed generation,'' {\em IET Smart Grid}, vol.~1, no.~3, pp.~85--95, 2018.

\bibitem{nois4}
A.~Hussain, C.-H. Kim, and S.~Admasie, ``An intelligent islanding detection of distribution networks with synchronous machine dg using ensemble learning and canonical methods,'' {\em IET Generation, Transmission \& Distribution}, vol.~15, no.~23, pp.~3242--3255, 2021.

\bibitem{nois5}
B.~K. Chaitanya, A.~Yadav, and M.~Pazoki, ``An advanced signal decomposition technique for islanding detection in dg system,'' {\em IEEE Systems Journal}, vol.~15, no.~3, pp.~3220--3229, 2020.

\bibitem{nois7}
A.~Hussain, C.-H. Kim, and M.~S. Jabbar, ``An intelligent deep convolutional neural networks-based islanding detection for multi-dg systems,'' {\em IEEE Access}, vol.~10, pp.~131920--131931, 2022.

\bibitem{pd1}
L.~Chen, X.~Dong, B.~Wang, L.~Shang, and C.~Liu, ``An edge computing-oriented islanding detection using differential entropy and multi-support vector machines,'' {\em IEEE Transactions on Smart Grid}, 2023.

\bibitem{nois6}
Y.~Xia, F.~Yu, X.~Xiong, Q.~Huang, and Q.~Zhou, ``A novel microgrid islanding detection algorithm based on a multi-feature improved lstm,'' {\em Energies}, vol.~15, no.~8, p.~2810, 2022.

\bibitem{(2)}
S.~B.~A. Bukhari, K.~K. Mehmood, A.~Wadood, and H.~Park, ``Intelligent islanding detection of microgrids using long short-term memory networks,'' {\em Energies}, vol.~14, no.~18, p.~5762, 2021.

\bibitem{vanwavenet}
A.~van~den Oord, S.~Dieleman, H.~Zen, K.~Simonyan, O.~Vinyals, A.~Graves, N.~Kalchbrenner, A.~Senior, and K.~Kavukcuoglu, ``Wavenet: A generative model for raw audio,'' in {\em 9th ISCA Speech Synthesis Workshop}, pp.~125--125, 2016.

\bibitem{myref}
A.~Alizadeh, S.~F. Zarei, and M.~Shateri, ``A sequence component-based feature for passive and artificial-intelligence-based islanding detection,'' 2024.

\bibitem{n04}
J.~C. Quispe and E.~Orduña, ``Transmission line protection challenges influenced by inverter-based resources: A review,'' {\em Protection and Control of Modern Power Systems}, vol.~7, pp.~1--17, July 2022.

\bibitem{n03}
Erdiwansyah, Mahidin, H.~Husin, Nasaruddin, M.~Zaki, and Muhibbuddin, ``A critical review of the integration of renewable energy sources with various technologies,'' {\em Protection and Control of Modern Power Systems}, vol.~6, no.~1, pp.~1--18, 2021.

\bibitem{Fselection}
P.~Dhal and C.~Azad, ``A comprehensive survey on feature selection in the various fields of machine learning,'' {\em Applied Intelligence}, vol.~52, no.~4, pp.~4543--4581, 2022.

\bibitem{cover2006elements}
T.~M. Cover and J.~A. Thomas, ``Elements of information theory, 2nd edition,'' {\em Willey-Interscience: NJ}, 2006.

\bibitem{bengio1994learning}
Y.~Bengio, P.~Simard, P.~Frasconi, {\em et~al.}, ``Learning long-term dependencies with gradient descent is difficult,'' {\em IEEE transactions on neural networks}, vol.~5, no.~2, pp.~157--166, 1994.

\bibitem{hochreiter1997long}
S.~Hochreiter and J.~Schmidhuber, ``Long short-term memory,'' {\em Neural computation}, vol.~9, no.~8, pp.~1735--1780, 1997.

\bibitem{gers1999learning}
F.~A. {Gers}, J.~{Schmidhuber}, and F.~{Cummins}, ``Learning to forget: continual prediction with lstm,'' in {\em 1999 Ninth International Conference on Artificial Neural Networks ICANN 99. (Conf. Publ. No. 470)}, vol.~2, pp.~850--855 vol.2, Sep. 1999.

\bibitem{he2016deep}
K.~He, X.~Zhang, S.~Ren, and J.~Sun, ``Deep residual learning for image recognition,'' in {\em Proceedings of the IEEE conference on computer vision and pattern recognition}, pp.~770--778, 2016.

\bibitem{ronneberger2015u}
O.~Ronneberger, P.~Fischer, and T.~Brox, ``U-net: Convolutional networks for biomedical image segmentation,'' in {\em Medical Image Computing and Computer-Assisted Intervention--MICCAI 2015: 18th International Conference, Munich, Germany, October 5-9, 2015, Proceedings, Part III 18}, pp.~234--241, Springer, 2015.

\bibitem{saito2015precision}
T.~Saito and M.~Rehmsmeier, ``The precision-recall plot is more informative than the roc plot when evaluating binary classifiers on imbalanced datasets,'' {\em PloS one}, vol.~10, no.~3, p.~e0118432, 2015.

\end{thebibliography}

\vspace{12pt}

\end{document}